\documentstyle[12pt]{article}
\newcommand{\Tr}{{\rm Tr}}
\newcommand{\sm}{\sum\limits_n\Tr[}
\newcommand{\yu}{+DG]+O(a^8)}
\newcommand{\pd}{D^2}
\newcommand{\pb}{B^2}
\newcommand{\oneone}{\setlength{\unitlength}{.2cm}\raisebox{-1cm}{
   \begin{picture}(1.2,1.2)(-1.8,-1.8)
   \put(0,0){\line(1,1){2}}
   \put(2,2){\line(1,0){4}}
   \put(6,2){\line(0,1){4}}
   \put(6,6){\line(-1,0){4}}
   \put(2,6){\line(-1,-1){2}}
   \put(0,4){\line(-1,0){4}}
   \put(-4,4){\line(0,-1){4}}
   \put(-4,0){\line(1,0){4}}
   \put(0,0){\circle*{.4}}
   \put(0,-.25){\makebox(0,0)[tr]{\footnotesize $an$}}
   \put(3,-1){\vector(1,0){2}}
   \put(3,-1){\vector(0,1){2}}
   \put(3,-1){\vector(1,1){2}}
   \put(5.7,-1){\makebox(0,0)[r]{\footnotesize $i$}}
   \put(2.7,0.5){\makebox(0,0)[r]{\footnotesize $k$}}
   \put(5.7,0.4){\makebox(0,0)[r]{\footnotesize $j$}}
   \end{picture}}}

\newcommand{\onetwo}{\setlength{\unitlength}{.4cm}\raisebox{-1.5cm}{
   \begin{picture}(1.2,1.2)(-1.8,-1.8)
   \put(0,0){\line(1,0){2}}
   \put(2,0){\line(0,1){2}}
   \put(2,2){\line(-1,0){2}}
   \put(0,2){\line(1,1){1}}
   \put(1,3){\line(-1,0){2}}
   \put(-1,3){\line(0,-1){2}}
   \put(-1,1){\line(1,0){2}}
   \put(1,1){\line(-1,-1){1}}
   \put(0,0){\circle*{.2}}
   \put(-.2,-.3){\makebox(0,0)[r]{\footnotesize $an$}}
   \end{picture}}}

\newcommand{\twoone}{\setlength{\unitlength}{.4cm}\raisebox{-1.5cm}{
   \begin{picture}(3.8,2.4)(-3,-3)
   \put(0,0){\line(0,1){2}}
   \put(0,2){\line(1,0){2}}
   \put(2,2){\line(1,1){1}}
   \put(3,3){\line(-1,0){2}}
   \put(1,3){\line(0,-1){2}}
   \put(1,1){\line(-1,0){2}}
   \put(-1,1){\line(-1,-1){1}}
   \put(-2,0){\line(1,0){2}}
   \put(0,0){\circle*{.2}}
   \put(0,-.2){\makebox(0,0)[tr]{\footnotesize $an$}}
   \put(1.5,-0.5){\vector(1,0){1}}
   \put(1.5,-.5){\vector(0,1){1}}
   \put(1.5,-.5){\vector(1,1){1}}
   \put(2.5,-1){\makebox(0,0)[r]{\footnotesize $i$}}
   \put(2.8,0.4){\makebox(0,0)[r]{\footnotesize $j$}}
   \put(1.2,0.5){\makebox(0,0)[r]{\footnotesize $k$}}
   \end{picture}}}

\newcommand{\twotwo}{\setlength{\unitlength}{.4cm}\raisebox{-1.5cm}{
   \begin{picture}(1.6,1.6)(-1.5,-3.2)
   \put(0,0){\line(1,0){2}}
   \put(2,0){\line(1,1){1}}
   \put(3,1){\line(-1,0){2}}
   \put(1,1){\line(0,1){2}}
   \put(1,3){\line(-1,0){2}}
   \put(-1,3){\line(-1,-1){1}}
   \put(-2,2){\line(1,0){2}}
   \put(0,2){\line(0,-1){2}}
   \put(0,0){\circle*{.2}}
   \put(0,-.2){\makebox(0,0)[r]{\footnotesize $an$}}
   \end{picture}}}

\newcommand{\threeone}{\setlength{\unitlength}{.15cm}\raisebox{-1cm}{
   \begin{picture}(3.8,2.4)(-6,-3.2)
   \put(-.5,0){\line(1,6){1}}
   \put(.5,6){\line(1,1){3}}
   \put(3.5,9){\line(0,-1){6}}
   \put(3.5,3){\line(-1,-1){3}}
   \put(.5,0){\line(-1,6){1}}
   \put(-.5,6){\line(-1,0){6}}
   \put(-6.5,6){\line(0,-1){6}}
   \put(-6.5,0){\line(1,0){6}}
   \put(0,0){\circle*{.8}}
   \put(0,-.8){\makebox(0,0)[tr]{\footnotesize $an$}}
   \put(4,-1){\vector(1,0){3}}
   \put(4,-1){\vector(0,1){3}}
   \put(4,-1){\vector(1,1){3}}
   \put(7,-2.2){\makebox(0,0)[r]{\footnotesize $i$}}
   \put(6.6,2){\makebox(0,0)[r]{\footnotesize $j$}}
   \put(3.5,1.2){\makebox(0,0)[r]{\footnotesize $k$}}
   \end{picture}}}

\newcommand{\threetwo}{\setlength{\unitlength}{.15cm}\raisebox{-1cm}{
   \begin{picture}(3.8,2.4)(-1,-3.2)
   \put(-.5,0){\line(1,6){1}}
   \put(.5,6){\line(1,1){3}}
   \put(3.5,9){\line(0,-1){6}}
   \put(3.5,3){\line(-1,-1){3}}
   \put(.5,0){\line(-1,6){1}}
   \put(-.5,6){\line(1,0){6}}
   \put(5.5,6){\line(0,-1){6}}
   \put(5.5,0){\line(-1,0){6}}
   \put(0,0){\circle*{.8}}
   \put(0,-.8){\makebox(0,0)[tr]{\footnotesize $an$}}
   \end{picture}}}

\newcommand{\threethree}{\setlength{\unitlength}{.15cm}\raisebox{-1cm}{
   \begin{picture}(3.8,2.4)(-6,-3.2)
   \put(-.5,0){\line(1,6){1}}
   \put(.5,6){\line(-1,-1){3}}
   \put(-2.5,3){\line(0,-1){6}}
   \put(-2.5,-3){\line(1,1){3}}
   \put(.5,0){\line(-1,6){1}}
   \put(-.5,6){\line(-1,0){6}}
   \put(-6.5,6){\line(0,-1){6}}
   \put(-6.5,0){\line(1,0){6}}
   \put(0,0){\circle*{.8}}
   \put(3,-.8){\makebox(0,0)[r]{\footnotesize $an$}}
   \end{picture}}}

\newcommand{\threefour}{\setlength{\unitlength}{.15cm}\raisebox{-1cm}{
   \begin{picture}(3.8,2.4)(-2,-3.2)
   \put(-.5,0){\line(1,6){1}}
   \put(.5,6){\line(-1,-1){3}}
   \put(-2.5,3){\line(0,-1){6}}
   \put(-2.5,-3){\line(1,1){3}}
   \put(.5,0){\line(-1,6){1}}
   \put(-.5,6){\line(1,0){6}}
   \put(5.5,6){\line(0,-1){6}}
   \put(5.5,0){\line(-1,0){6}}
   \put(0,0){\circle*{.8}}
   \put(2.5,-.8){\makebox(0,0)[tr]{\footnotesize $an$}}
   \end{picture}}}

\newcommand{\ftwo}{\setlength{\unitlength}{.15cm}\raisebox{-.8cm}{
   \begin{picture}(3.8,2.4)(-1,-3.2)
   \put(0,-1){\line(3,1){6}}
   \put(6,1){\line(0,1){6}}
   \put(6,7){\line(-1,0){6}}
   \put(0,7){\line(0,-1){6}}
   \put(0,1){\line(3,-1){6}}
   \put(6,-1){\line(0,-1){6}}
   \put(6,-7){\line(-1,0){6}}
   \put(0,-7){\line(0,1){6}}
   \put(0,0){\circle*{.8}}
   \put(-1,0){\makebox(0,0)[r]{\footnotesize $an$}}
   \put(8.5,-2){\vector(1,0){3}}
   \put(8.5,-2){\vector(0,1){3}}
   \put(11.5,-3){\makebox(0,0)[r]{\footnotesize $i$}}
   \put(8,2){\makebox(0,0)[r]{\footnotesize $j$}}
   \end{picture}}}

\newcommand{\fone}{\setlength{\unitlength}{.15cm}\raisebox{-1cm}{
   \begin{picture}(3.8,2.4)(-6,-3.2)
   \put(-.5,0){\line(1,6){1}}
   \put(.5,6){\line(1,0){6}}
   \put(6.5,6){\line(0,-1){6}}
   \put(6.5,0){\line(-1,0){6}}
   \put(.5,0){\line(-1,6){1}}
   \put(-.5,6){\line(-1,0){6}}
   \put(-6.5,6){\line(0,-1){6}}
   \put(-6.5,0){\line(6,0){6}}
   \put(0,0){\circle*{.8}}
   \put(0,-1.2){\makebox(0,0)[r]{\footnotesize $an$}}
   \end{picture}}}

\newcommand{\five}{\setlength{\unitlength}{.4cm}\raisebox{-.8cm}{
   \begin{picture}(3.8,2.4)(-.8,-.8)
   \put(0,0){\line(0,1){2}}
   \put(0,2){\line(1,0){2}}
   \put(2,2){\line(0,-1){2}}
   \put(2,0){\line(-1,0){2}}
   \put(0.2,0.2){\line(0,1){2}}
   \put(0.2,2.2){\line(1,0){2}}
   \put(2.2,2.2){\line(0,-1){2}}
   \put(2.2,0.2){\line(-1,0){2}}
   \put(0,0){\circle*{.2}}
   \put(-.1,-.1){\makebox(0,0)[r]{\footnotesize $an$}}
   \put(3,-0.5){\vector(1,0){1.5}}
   \put(3,-.5){\vector(0,1){1.5}}
   \put(4.5,-1){\makebox(0,0)[r]{\footnotesize $i$}}
   \put(2.8,1){\makebox(0,0)[r]{\footnotesize $j$}}
   \end{picture}}}

\newcommand{\plaq}{\setlength{\unitlength}{.4cm}\raisebox{-.7cm}{
   \begin{picture}(1.4,1.4)(-1,-1)
   \put(0,0){\line(0,1){2}}
   \put(0,2){\line(1,0){2}}
   \put(2,2){\line(0,-1){2}}
   \put(2,0){\line(-1,0){2}}
   \put(0,0){\circle*{.2}}
   \put(0,-.2){\makebox(0,0)[tr]{\footnotesize $an$}}
   \put(3,-0.5){\vector(1,0){1.5}}
   \put(3,-.5){\vector(0,1){1.5}}
   \put(4.5,0){\makebox(0,0)[r]{\footnotesize $i$}}
   \put(2.8,1){\makebox(0,0)[r]{\footnotesize $j$}}
   \end{picture}}}

\begin{document}
\title{The First Calculation for the Mass of the Ground $4^{++}$ Glueball
State on Lattice}

\author{{Da Qing Liu$^2$, Ji Min Wu$^{1,~2}$}\\
        {\small $^1$CCAST(World Lab. ), P. O. Pox 8730, Beijing 100080,
                China}\\
        {\small $^2$Institute of High Energy Physics, Chinese Academy
                of
 Sciences,}\\
        {\small P. O. Box 918-4, Beijing, 100039, P. R. China}
       }
\maketitle

\begin{center}
\begin{minipage}{5.5in}
\vskip 0.8in
{\bf Abstract}\\
Under the quenched approximation, we perform a lattice calculation for
the mass of the ground $4^{++}$ glueball state in $E^{++}$ channel
on a $D=3+1$ lattice. Our calculation shows that the
mass of this state is $M_G(4^{++})=3.65(6)(18)GeV$, which rules out the
$4^{++}$ or mainly $4^{++}$ glueball state
interpretation for $\xi(2230)$.

\end{minipage}
\end{center}
\vskip 1in
\indent

\newpage
\section {Introduction}

As the experiments reported$^{\cite{exprm}}$, there maybe exist
one candidate for glueball state, namely, $\xi(2230)$, which
has been seen mainly in $J/\Psi$ radiative decay with a
number of decay channels. But its spin has not been
determined  experimentally:$^{\cite{exprm1}}$ we don't know whether
its spin is $2^{++}$ or $4^{++}$.
Therefore, one hopes to calculate the masses of $2^{++}$ and $4^{++}$
glueball states on lattice for the spin interpretation of $\xi(2230)$
since lattice QCD is regarded as the most reliable theory to study hadron
spectra.

Meanwhile, lattice simulation has achieved tremendous success in the study
of glueball states. For instance, some authors show mass
spectra$^{\cite{sim,morning}}$ for glueball states. But, so far we do not know how
to calculate the mass of $4^{++}$ glueball state assuredly on $D=3+1$
lattice. On a $D=2+1$ cubic lattice, in Ref. \cite{tepper}, Johnson and
Teper performed a
calculation for these states. Nevertheless, there are some disputes
about the interpretation of the calculation, i.e., one does not know
whether the mass calculated in Ref. \cite{tepper} is really the mass
of $4^{++}$ glueball states. In addition, as authors consider,
 one can hardly extent the construction of the operator in Ref.
\cite{tepper} into the $D=3+1$ lattice.

Therefore, from the point of views of both experiment and theory,
it is important to calculate the $4^{++}$ glueball mass on lattice.
Basing on the connection between the
continuum limits of the  asymptotic expansion of the chosen operator and
the construction of the operator with definite $J^{PC}$, we have developed a
new procedure$^{ \cite{liudq}}$
which can calculate masses of glueball states with arbitrary
$J^{PC}$. According to this procedure,
a lattice calculation for the mass of the ground $4^{++}$
glueball states is performed in this paper by the construction of
operators, corresponding states of which are in $E^{++}$ channel in the
simulation. The leading term in expansion of this operator belongs to
$4^{++}$ representation of $SO(3)^{PC}$  group. Therefore, from the
asymptotic expansion of this constructed operators,
we assure that, in continuum case, the spin of corresponding states
is $4^{++}$. Then, the mass of $4^{++}$ glueball state was obtained.
We found that the calculated mass is prominently different with
that of $2^{++}$ glueball states $^{ \cite{liudq}}$
and it is about 1.5 and 2 times  of that of tensor and scalar
glueball state respectively. We should  argued  that our results rule out
the possible interpretation of the ground $4^{++}$ glueball states for
$\xi(2230)$.

This paper is organized as follows. In section 2 we show how to
construct $4^{++}$ glueball operators in $E^{++}$ channel. We
should also present our results and conclusions in section 3 and section 4.

\section{ The Construction of the Operator Corresponding to $4^{++}$
Glueball state}

As we know, a definite state $|\psi>$ is generated by a current
$o$ acting on vacuum $|0>$:
\begin{equation} \label{e1}
|\psi>=o|0>,
\end{equation}
and the character of $|\psi>$ can be described by the current $o$.
Therefore, as mentioned in \cite{liudq}, to calculate the
mass of glueball state with spin $J^{PC}=4^{++}$, we should carry
out three steps. First, we should
write out nonet currents
which transform as the representation $4^{++}$ under the $SO(3)^{PC}$
group and decompose them into irreducible representations
$A_1^{++}$,
$E^{++}$, $T_1^{++}$ and $T_2^{++}$ of the group $O^{PC}$ ( the cubic
group with parity and charge conjugation transform ) in the
subduced representation \footnote { The subduced representation used
here is the representation obtained by trivially embedding the
$O^{PC}$ group into the $SO(3)^{PC}$ group.}. Then we construct
corresponding operators on lattice. Its continuum limits
should be just those currents mentioned above. Next
we should also use the variantional procedure to extract masses
of the ground $4^{++}$ states and excited $4^{++}$ states.

We shall calculate the mass of the ground $4^{++}$
glueball states in this paper.
Among nonet currents, there is only one type of such nonet current,
the mass dimension of which is the lowest one. This
nonet current is of the form
$Tr(\vec{B}\vec{D}\vec{D}\vec{B})$ and we expect that it
should give the most contribution
to the ground states with spin $J^{PC}=4^{++}$ due to dimensional
analysis. Therefore, we will only consider such nonet current
and the construction of corresponding operators on lattice in this paper.
Apparently such disposal will bring up some errors and its affectation
will be considered in the error estimate in section 3.

\paragraph {The Nonet Current and Its Decomposition}

To simplify, we denote magnetic field $B_{x(y,~z)}=
B_{x(y,~z)}^a{\lambda^a \over 2}$ by $B_{1(2,~3)}$
and covariant derivative $D_{x(y,~z)}$ by
$D_{1(2,~3)}$ ($D_i\cdot=\partial_i-i[A_i,\cdot]$) respectively here.
Obviously one can treat $B_i$'s as bases of the spin $J=1$,
and  $B_{\pm}=\mp{1\over \sqrt{2}}(B_1\pm
iB_2)$ and $B_0=B_3$ as three standard bases\footnote {Here $B_{\pm}$ and
$B_0$ is the eigen-current of the operator $\hat{J}_3$,
i.e., $\hat{J}_3(B_{\pm})=\pm B_{\pm}$ and $\hat{J}_3(B_0)=0$.}.
It is similar for covariant derivatives $D_i$'s, which can also
be combined into three standard bases: $D_{\pm}=\mp{1\over \sqrt{2}}(D_1\pm
iD_2)$ and $D_0=D_3$.

Since spins of magnetic field $\vec{B}$ and covariant
derivative operator $\vec{D}$ are both 1, the nonzero gauge invariant
nonet current of the mass dimension 6 is sole, so that
according to C-G coefficients, one can write these 9 gauge
invariant currents unambiguously in the standard form up to full covariant
derivatives (The following $c_i$'s$(i=4,\cdots,-4)$ are eigen-currents of
operator $\hat{J}_3$, i.e., $\hat{J}_3(c_i)=i~c_i$.)\footnote{
Phenomenologically, we can regard this nonet current as
 $^5D_4$ multi-current and the corresponding state as $^5D_4$
multi-state.}:
\begin{eqnarray}
\label{a4}
c_4&=&\Tr [B_+D_+D_+B_++DG], \\
\label {a3}
  c_3&=&\Tr [B_+D_+D_0B_++B_+D_+D_+B_0+DG], \\
\label{a2}
c_2&=&{1\over \sqrt{7}}\Tr [(B_+D_+D_-B_++B_+D_0D_0B_++B_+D_+D_+B_-
\nonumber \\
&&+B_0D_+D_+B_0+2B_+D_+D_0B_0+2B_+D_0D_+B_0)], \\
  c_1&=&{1\over \sqrt{7}}\Tr [B_+D_0D_-B_++B_0D_+D_+B_-+B_+D_+D_-B_0
  \nonumber \\
 &&B_+D_-D_+B_0+2B_+D_0D_0B_0+B_+D_+D_0B_-
 \nonumber \\
 \label{a1}
 &&+B_+D_0D_+B_-+2B_0D_+D_0B_0 + DG], \\
 c_0&=&{1\over \sqrt{70}}\Tr [B_+D_-D_-B_+ + B_-D_+D_+B_-]
 \nonumber \\
&&+{2\over \sqrt{70}}\Tr [B_+D_+D_-B_- + B_+D_-D_+B_-]
\nonumber \\
 &&+{4\over \sqrt{70}}\Tr [B_+D_-D_0B_0 + B_+D_0D_-B_0
 \nonumber \\
  &&+B_-D_+D_0B_0+ B_0D_+D_0B_-
  \nonumber \\
  \label{a0}
  &&+B_+D_0D_0B_- + B_0D_+D_-B_0+B_0D_0D_0B_0 +DG],\\
  c_{-1}&=&{1\over \sqrt{7}}\Tr [B_-D_0D_+B_- +B_0D_-D_-B_+ + B_-D_-D_+B_0
  \nonumber \\
 &&B_-D_+D_-B_0 + 2B_-D_0D_0B_0 + B_-D_-D_0B_+
 \nonumber \\
 \label{a-1}
 &&+B_-D_0D_-B_+ + 2B_0D_-D_0B_0 +DG],\\
\label{a-2}
c_{-2}&=&{1\over \sqrt{7}}\Tr [(B_-D_-D_+B_- + B_-D_0D_0B_- +
B_-D_-D_-B_+
\nonumber \\
&&+ B_0D_-D_-B_0 + 2B_-D_-D_0B_0 + 2B_-D_0D_-B_0) +DG], \\
\label{a-3}
  c_{-3}&=&\Tr [B_-D_-D_0B_- + B_-D_-D_-B_0 +DG], \\
\label{a-4}
c_{-4}&=&\Tr [B_-D_-D_-B_- +DG].
\end{eqnarray}
Here $DG$ is some full covariant derivative terms. In fact,
one can get $c_{-i}(i=1,~2,~3,~4)$ easily by the swap $+ \leftrightarrow
-$ in $c_i$.

Then we turn to the $O^{PC}$ group, the finite
subgroup of $SO(3)^{PC}$ group, to form the bases of subduced
representation with nonet current mentioned above.
 As shown in Ref. \cite{berg}, the subduced representation is reducible
 and one can reduce this representation into irreducible ones $A_1^{++},~E^{++},
 ~T_1^{++}$ and $T_2^{++}$, the dimension of which are 1,~2,~3,~3
 respectively. For our aim we only show the result of the decomposition
 into $E^{++}$ here.

According to the procedure introduced in Ref. \cite{liu2},
we can get two bases of the representation $E^{++}$
\begin{eqnarray}
e_1&=&-{1\over \sqrt{2}}(c_2+c_{-2})
\nonumber \\
& \propto & \Tr [B_1(D_1^2 -D_3^2) B_1 -B_2(D_2^2-D_3^2)B_2
+B_3(D_2^2-D_1^2)B_3
\nonumber \\
&& + 4B_2D_3D_2B_3 - 4B_1D_3D_1B_3 +DG];
 \nonumber \\
e_2&=&{1\over 2}(\sqrt{{7\over 6}}c_4+\sqrt{{7\over 6}}c_{-4}-\sqrt{{5\over
3}}c_0)
\nonumber \\
&\propto & \Tr [B_1(D_1^2+D_3^2-2D_2^2)B_1
+ B_2(D_2^2+D_3^2-2D_1^2)B_2
\nonumber \\
&&+ B_3(D_1^2+D_2^2-2D_3^2)B_3 + 4B_1D_3D_1B_3
\nonumber \\
\label{e-2}
&&+ 4B_2D_3D_2B_3 -8B_1D_2D_1B_2 + DG].
\end{eqnarray}

These two currents are combinations of following three linear dependent ones:

\begin{eqnarray}
e^{(1)}&=&\Tr [B_2(D_2^2 -D_1^2) B_2 -B_3(D_3^2-D_1^2)B_3
+B_1(D_3^2-D_2^2)B_1
\nonumber \\
&& + 4B_3D_1D_3B_1 - 4B_2D_1D_2B_1 +DG];
\nonumber \\
e^{(2)}&=&\Tr [B_3(D_3^2 -D_2^2) B_3 -B_1(D_1^2-D_2^2)B_1
+B_2(D_1^2-D_3^2)B_2
\nonumber \\
\label{e13}
&& + 4B_1D_2D_1B_2 - 4B_3D_2D_3B_2 +DG];
\nonumber \\
e^{(3)}&=&\Tr [B_1(D_1^2 -D_3^2) B_1 -B_2(D_2^2-D_3^2)B_2
+B_3(D_2^2-D_1^2)B_3
\nonumber \\
\label{current}
&& + 4B_2D_3D_2B_3 - 4B_1D_3D_1B_3 +DG].
\end{eqnarray}

These three currents are of equivalence up to a cyclic
permutation $1 \rightarrow 2 \rightarrow 3 \rightarrow 1$
of superscripts and subscripts in Eq. (\ref{current}).

Then we consider how to construct operators on lattice, which are in
the representation $E^{++}$ and their continuum limits  are
just $e^{(i)}(i=1,~2,~3)$.

\paragraph {The Construction of Operators on Lattice}
In this paragraph, we assign that $(i,~j,~k)$ is even permutation of
$(1,~2,~3)$, i.e., $(i,~j,~k)=(1,~2,~3),~(2,~3,~1)$ or $(3,~1,~2)$.

Using the link variable as defined in the discussion of improved
action$^{\cite{s11,s12}}$:
\begin{equation}
U(n,i)=T\exp( \int^a_0 dt A_i(an+\hat{i}t),
\end{equation}
 we denote Wilson operators by figures
\footnote{We use a notation, where $A_i$ is an anti-hermitian,
traceless $N\times N$-matrix.}. For instance,
we denote the plaquette operator
 \begin{equation} O_{ij}=\sum\limits_n
O_{ij}(n)=Re \sum\limits_n
\Tr [U(n,i)U(n+\widehat{i},j)U^{-1}(n+\widehat{j},i)U^{-1}(n,j)]
\end{equation}
by the figure $\sum\limits_n(\plaq\mbox{}\hspace{0.7in})$ for simplification.
Here the summation is only over  all the spatial links in the same time slice.

Let's consider the following five types of operators  $d^{(1)}_k -
d^{(5)}_k$:
\begin{description}
\item[1.]        ~~
The first type of operators are $d^{(1)}_k(k=1,~2,~3)$. They are from the
following operators:
\begin{equation}
b_k^{(1)}=\epsilon_{ijk}^{(1)}+\epsilon_{jik}^{(1)},
\end{equation}
where
\begin{equation}
\epsilon_{ijk}^{(1)}=\sum\limits_{n}(\mbox{}\hspace{0.2in}\oneone
\mbox{}\hspace{0.6in}+\onetwo \mbox{}\hspace{0.5in}).
\end{equation}

The expansion of the operator $b_k^{(1)}$ according to lattice spacing $a$
is
\begin{eqnarray}
b_k^{(1)}&=&\sm {\bf 4}+5a^4(B_j^2+B_i^2)+{a^6\over 12}B_i
          (17D_j^2+5D_k^2+12D_i^2)B_i
\nonumber \\
&&+{a^6 \over 12}B_j(17\pd_i+5\pd_k+12\pd_j)B_j-2a^6B_iD_jD_iB_j
\nonumber \\
&&\yu .
\end{eqnarray}
Here $B_i=-{1\over 2}\sum\limits_{jk}\epsilon_{ijk}F_{jk}$ and
$F_{ij}=\partial_i A_j-\partial_j A_i+[A_i,A_j]$.

Then the operator $d_k^{(1)}$ is:
\begin{eqnarray}
d_k^{(1)}&=&b^{(1)}_i-b^{(1)}_j
\nonumber \\
&=&\sm 5a^4(\pb_j-\pb_i)+{a^6\over 12}B_j(17\pd_k+5\pd_i+12\pd_j)B_2
\nonumber \\
&&-{a^6\over 12}B_i(17\pd_k+5\pd_j+12\pd_i)B_i+a^6B_k(\pd_j-\pd_i)B_k
\nonumber \\
\label{op1}
&&-2a^6(B_jD_kD_jB_k-B_iD_kD_iB_k)\yu.
\end{eqnarray}

\item[2.]      ~~
Next, we consider operators $d^{(2)}_k(k=1,~2,~3)$.  They are from
operators
\begin{equation}
b_k^{(2)}=\epsilon_{ijk}^{(2)}+\epsilon_{jik}^{(2)},
\end{equation}
where
\begin{equation}
\epsilon_{ijk}^{(2)}=\sum\limits_{x}(\twoone
\mbox{}\hspace{0.6in}+\twotwo \mbox{}\hspace{0.5in}).
\end{equation}
The definition and the expansion according to lattice spacing $a$
of operators $d^{(2)}_k$ are
\begin{eqnarray}
d_k^{(2)}&=&b^{(2)}_i-b^{(2)}_j
\nonumber \\
&=&\sm 7a^4(\pb_i-\pb_j)+{a^6\over 12}B_i(19\pd_j+19\pd_k+24\pd_i)B_i
\nonumber \\
&&-{a^6\over 12}B_j(19\pd_i+19\pd_k+24\pd_j)+a^6(B_jD_kD_jB_k-
B_iD_kD_iB_k)
\nonumber \\
&&\yu.
\end{eqnarray}

\item[3.]    ~~
Then we study operators $d^{(3)}_k$, which are from operators
\vskip .1in
\begin{eqnarray}
b^{(3)}_k&=&\epsilon_{ijk}^{(3)}
\nonumber \\
&=&\sum\limits_{x}(\threeone \mbox{}\hspace{0.6in}+\threetwo\mbox{}
\hspace{0.3in} +\threethree\mbox{}\hspace{0.3in} +\threefour
\mbox{}\hspace{0.3in}).
\end{eqnarray}
For $d^{(3)}_k$ we get
\begin{eqnarray}
d_k^{(3)}&=&b^{(3)}_i-b^{(3)}_j
\nonumber \\
&=&\sm 2a^4(\pb_j-\pb_i)+{a^6\over 6}B_j(\pd_i+\pd_k)B_j
-{a^6\over 6}B_i(\pd_j+\pd_k)B_i
\nonumber \\
&&+a^6(B_jD_kD_2B_k-B_iD_kD_iB_k)\yu.
\end{eqnarray}

\item[4.]  ~~
Operators $d^{(4)}_k$ are from
\begin{eqnarray}
vb^{(4)}_k&=&\epsilon_{ij}^{(4)}
\nonumber \\
&=&\sum\limits_{x}(\fone \mbox{}\hspace{0.6in}+\ftwo \mbox{}\hspace{0.6in}).
\end{eqnarray}
And again
\begin{eqnarray}
d_k^{(4)}&=&b^{(4)}_i-b^{(4)}_j
\nonumber \\
&=&\sm {a^6\over 2}(B_j(\pd_i+\pd_k)B_j-B_i(\pd_j+\pd_k)B_i
\nonumber \\
&&~~~~\yu.
\end{eqnarray}

\item[5.]          ~~
At last we study the operator $d^{(5)}_k$ which are constructed by
\begin{eqnarray}
b^{(5)}_{k}=\epsilon^{(5)}_{ijk}=\sum\limits_n(\five \mbox{}\hspace{0.2in}).
\end{eqnarray}
The construction is
\begin{eqnarray}
d_k^{(5)}&=&b^{(5)}_i-b^{(5)}_j
\nonumber \\
&=&\sm 2a^4(\pb_i-\pb_j)+{a^6\over 6}B_i(\pd_j+\pd_k)B_i
\nonumber \\
&& -{a^6\over 6}B_j(\pd_i+\pd_k)B_j \yu.
\end{eqnarray}

\end{description}

Therefore we let
\begin{eqnarray}
e^k&=&d^{(1)}_k+d^{(2)}_k+5d^{(3)}_k+2d^{(4)}_k+4d^{(5)}_k
\nonumber \\
&=&a^6\sm B_i(D_i^2 -D_k^2) B_i -B_j(D_j^2-D_k^2)B_j
+B_k(D_j^2-D_i^2)B_k
\nonumber \\
\label{current1}
&& + 4B_jD_kD_jB_k - 4B_iD_kD_iB_k \yu.
\end{eqnarray}

While comparing Eq. (\ref{current1}) and Eq. (\ref{current}), we
find that operators $e^k(k=1,~2,~3)$ are just our aimed operators.
Noticing that these operators are all 8-link,
so that we need not consider the "tadpole" renormalization due
to the mean field theory$^{\cite{ren}}$. We should utilize operators $e^k$ to
calculate the mass of $4^{++}$ glueball states in this paper.

Results and some discussions are shown in the forthcoming section.

\section{Calculation Results}
Under the quench approximation, we perform our calculation on
an anisotropic $8^3\times 40$ lattice with improved action
introduced in Ref. \cite{improve}:
\begin{equation}
S_{II}=\beta \{ {5\Omega_{sp}\over 3\xi u_s^4}+{4\xi\Omega_{tp}\over 3u_s^2u_t^2}
-{\Omega_{sr}\over 12 \xi u_s^6}-{\xi\Omega_{str}\over 12u_s^4u_t^2} \},
\end{equation}
where $\beta=6/g^2,~g$ is the QCD couple constant, $u_s$ and $u_t$ are mean link
renormalization parameters(we set $u_t=1$), $\xi=a_s/a_t=5.0$ is the aspect
ratio,  and $\Omega_{sp}$ includes the summation over all spatial plaquettes
on the  lattice, $\Omega_{tp}$ indicates the temporal plaquettes,
$\Omega_{sr}$  denotes the planar $2\times 1$ spatial rectangular loops and
$\Omega_{str}$  refers to the short temporal rectangles( one temporal and two
spatial links).  More detail is given in Ref. \cite{improve}. In each $\beta$
calculation, we set 2880 sweeps to make configurations reach to equilibrium
and then make   total 8100 measurements in 90 bins ( We perform measurement
one time every four sweeps). The other calculation parameters, such as
$u_s^4$, $r_s/r_0$ and  $a_s$, are as the same as those in Ref. \cite{morning}.

Our results in different  $\beta$ are shown in Table 1:
\begin{center}
\begin{tabular}{|c|c|c|c|c|c|} \hline
 $ \beta$   & 1.7       &1.9      &  2.2   &  2.4     &    2.5  \\ \hline
$4^{++}$    &1.61(1)    &1.35(7)  &0.97(4) &0.79(3)   & 0.70(5) \\ \hline
$2^{++}$    &1.019(3)   &0.95(1)  &0.71(2) &0.548(6)  &0.519(4)   \\ \hline
\end{tabular} \\
{\small {\bf Table. 1}~ $4^{++}$ glueball energy $ m_{G} a_t$ for each
$\beta$.  The numerals in the brackets are \\error estimates( The data for
$2^{++}$ glueball is from Ref. \cite{liudq}.).
 }
\end{center}

Now we discuss a little about the error estimate.

As discussed in many papers, one may regard that the chosen improved
action
breaks the rotation symmetry up to $O(a^4)$ in error estimate. But since
the mass dimension of terms which break the rotation symmetry up to $O(a^4)$
is eight and the lowest mass dimension of our chosen operators
in Eq. (\ref{current1}) is six, the up limit of precision here is $O(a^2)$.

On the other hand, among the currents with mass dimension 7 or 9, there is
no current with
 $P=+$. Therefore, the precision calculated here is actually
$O(a^2)$ and the systematic error can be written out as the form
$c_2a_s^2+c_4a_s^4+\cdots$, which should be used to fit our simulation data.
Simulation results and fitting curve are shown in Fig. 1. The bottom curve and
data$^{\cite{liudq}}$ in Fig. 1 are for the $2^{++}$ glueball state
which
also belong to $E^{++}$ channel.

From the data and fitting curve, the statistical error of the mass measurement
is $0.044GeV$. As argued in ref. \cite{morning}, there are 1\% systematic
error dur to aspect ratio. In addition, our approach leads into about
0.5\% systematic error.
Therefore, the total systematic error is about 1.1\%( $0.041 GeV$). So 
the mass of $4^{++}$ glueball states is $3.65(6)GeV$. Including
the uncertainty in $r_0^{-1}=410(20)Mev$, our final results are:
$M_G(4^{++})=3.65(6)(18)GeV$ which is about 1.5 times and twice of masses
of the tensor and scalar glueball states respectively.

\section {Conclusion}
Through the construction of operators in $E^{++}$ channel which are
complicated to some degree, we perform a calculation of mass of
the ground $4^{++}$ glueball state under the quenched approximation.
Due to the expansion of the chosen operator, we confirm that, in
continuum, the mass calculated here is the mass of the ground $4^{++}$
glueball state and it is $M_G(4^{++})=3.65(6)(18)GeV$.

Apparently, our result rules out the $J^{PC}=4^{++}$ glueball
interpretation
for $\xi(2230)$ even by taking into account the opinion that
the mass will be shifted by as much as 20\% in going to full
QCD$^{\cite{wig}}$. While noticing that the common calculated mass
for tensor glueball state is about $2.4GeV$ under the quench approximation,
we claim that our calculation supports the $2^{++}$ glueball interpretation
for $\xi(2230)$  if this state exists and it is confirmed as a glueball.

\newpage
\section*{Figure caption}
{\bf Figure 1}~~~~The mass of $4^{++}$ glueball against the lattice
spacing  $(a_s/r_0)^2$. 
Our fitting curve for the data of $4^{++}$ glueball state is 
$m(4^{++},a_s)=3.65-1.22(a_s/r_0)^2+2.74(a_s/r_0)^4(unit:GeV)$.
In continuum case, the mass of $4^{++}$ glueball state is 
$3.65(4)GeV$ if we only consider the statistical error.
The bottom data and curve:
$m(2^{++},a_s)=2.417+0.783(a_s/r_0)^2-0.787(a_s/r_0)^4 (unit:GeV)$ 
are for the mass of $2^{++}$ glueball state which is shown in 
Ref. \cite{liudq}.

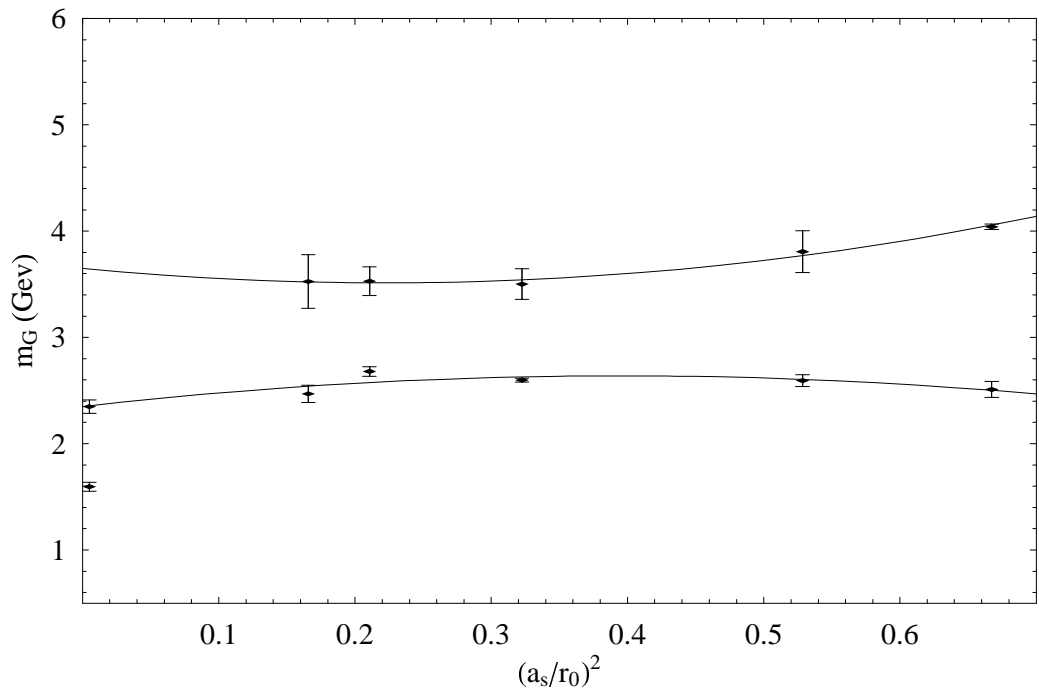
\end{document}